\documentclass[twocolumn,preprintnumbers,amsmath,amssymb]{revtex4}
\usepackage{tabularx,graphicx}

\usepackage{color}
\usepackage{hyperref}
\hypersetup{
    colorlinks=true,
    linkcolor=blue,
    filecolor=blue,      
    urlcolor=blue,
}
\usepackage{lipsum, babel}
\usepackage{color}

\usepackage{ulem}   % to strike things out
\begin{document}

%\documentstyle[aps]{revtex}
%\documentstyle[preprint,aps]{revtex}
%\begin{document}

\newcommand{\beq}{\begin{equation}}
\newcommand{\eeq}{\end{equation}}
\newcommand{\beqn}{\begin{eqnarray}}
\newcommand{\eeqn}{\end{eqnarray}}
\newcommand{\bmath}{\begin{subequations}}
\newcommand{\emath}{\end{subequations}}
\newcommand{\bra}[1]{\langle #1|}
\newcommand{\ket}[1]{|#1\rangle}

%\draft
\title{Hydride superconductivity:  here to stay, or to lead astray and  soon go away?}

\author{J. E. Hirsch}
\address{Department of Physics, University of California, San Diego,
La Jolla, CA 92093-0319}
 \begin{abstract} 

{In a recent Comment \cite{stay}, fifteen prominent leaders in the field of condensed matter physics   declare  that hydride superconductivity is real,
and urge funding agencies to continue to support the field.  I question the validity and constructiveness of their argument.}
 \end{abstract}
 \date{December 31, 2024}
\maketitle

        \begin{figure} [t]
 \resizebox{8.5cm}{!}{\includegraphics[width=6cm]{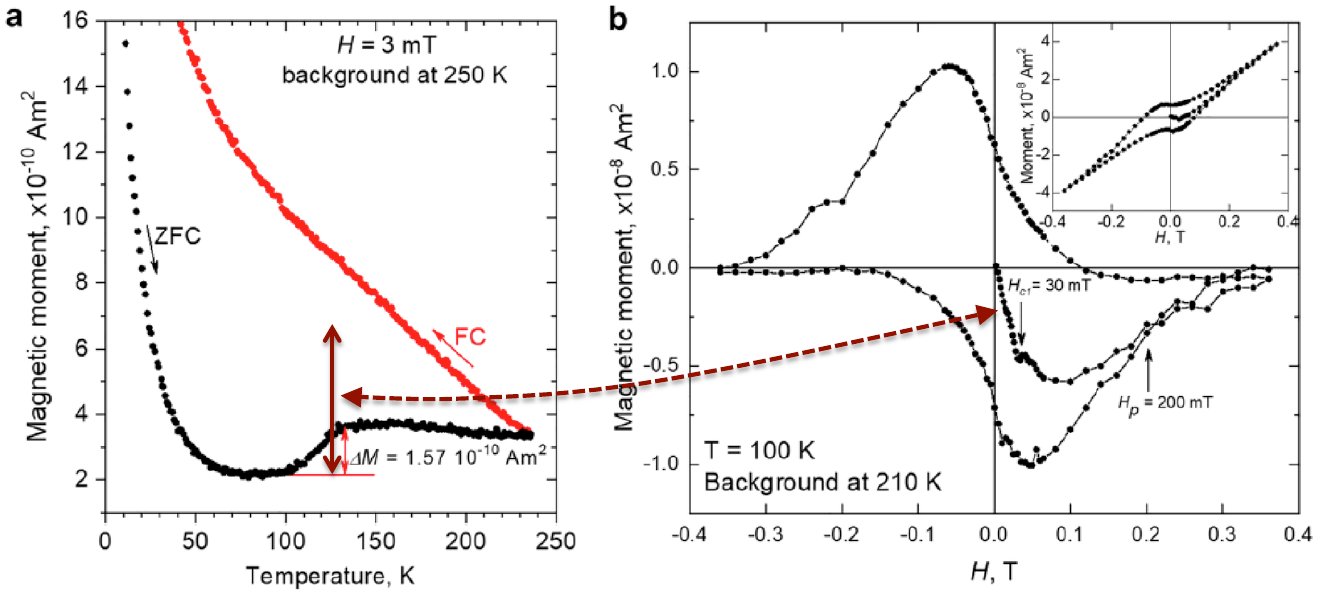}} 
 \caption {Fig. 6 of Ref. \cite{e2022p}. Its caption reads \cite{e2022p}:
 {\it ``Magnetization measurements of superconducting $H_3S$
at pressure $P \sim 140 GPa$. {\bf a} M(T) data measured at H = 3 mT after subtraction
of the background collected at 250 K. {\bf b} M(H) data measured at
100 K showing the initial virgin curve and hysteretic loop on alternating
the magnetic field. Inset shows the original hysteretic loop before
subtraction of the background signal originating from the DAC at
210 K. } Brown dashed line with double arrows and vertical line with double arrows were added here.}
 \label{figure1}
 \end{figure}

Reference \cite{stay} acknowledges that the field of superconductivity in hydrides under high pressure has been mired in   controversy, yet does not explain nor reference what the controversy is \cite{controv}. It discusses some selected experimental evidence that 
the authors  argue supports the validity of the claims of superconductivity in hydrides under high pressure, and concludes that 
{\it ``it is overwhelmingly probable that the superconductivity in this class of compounds is genuine''} \cite{stay}. 
Here I argue that it is overwhelmingly probable that they are wrong.

The authors of Ref. \cite{stay} are fifteen distinguished scientists
from renowned institutions across three continents with a strong record of scientific accomplishment, but none of them has worked on
hydrides under high pressure. Their opinion \cite{stay}  is not validated by a careful analysis of the data they are opining on, and as a consequence has no claim to scientific validity.  It can be characterized instead as a political manifesto by influential members of the physics community  that I argue is not constructive.

%
%with essentially zero track record in the field of hydride superconductivity. Yet they feel entitled to present, 
%instead of a balanced scientific analysis where both (or multiple) sides of a scientific  argument are discussed, a political manifesto with the following underlying content:  we have the expertise   to decide who is right and who is wrong, and we have  the moral authority  to urge funding agencies to continue devoting taxpayer's dollars to what we think is right, and to encourage young physicists to enter this field. 
% 

According to M. I. Eremets himself, the lead author of the 2015 paper \cite{e2015}  that started the hydride superconductivity
saga, the 2015 paper would never have been published without magnetic evidence. Quoting verbatim from Ref.  \cite{ehistory}: {\it ``Without these
measurements, Science and Nature rejected our paper even if we had already substantial
shreds of evidence of superconductivity (zero resistance, proper shift of $T_c$ in a magnetic
field, and isotope effect). We needed a quick solution.''}  The ``quick solution'' was the measurements shown in Figs. 1{\bf d} and 1{\bf e}
of the manifesto Ref. \cite{stay},  1{\bf d} published in 2015 and 1{\bf e} measured in 2015 but only seeing the light of day
in 2022 \cite{e2022p}, for unexplained reasons.

Fig. 1{\bf e}  of Ref. \cite{stay} is shown  here  as Fig. 1{\bf b}.   Ref. \cite{stay} describes this figure as follows:
{\it ``... shows the loop reported in Ref. 6.} (Ref. \cite{e2022p} here). {\it Data including
the background are shown in the inset. Even before background subtraction, the data for the virgin curve go negative,
in low applied fields, offering strong evidence that diamagnetism is observed in the raw data. The simple subtraction
of the linear background gives the curve shown in the main plot, which has the main qualitative features expected of
a superconducting hysteresis loop in the presence of flux trapping.''}

The left panel of Fig. 1, Fig. 1{\bf a}, also from Ref. \cite{e2022p}, shows
magnetic moment versus temperature for the same sample under the same pressure  for which the right panel shows the hysteresis loop.
There are several peculiar features of Fig 1{\bf a}: (i) the curves for ZFC and FC diverge at temperature $T\sim 240K$, well above
the assumed critical temperature $T\sim 140K$. That is $not$ the behavior expected for a superconducting sample, for which  the curves should start diverging when the temperature is lowered
right below $T_c$, rather than $100K$ above $T_c$. (ii) Given the magnitude of the diamagnetic moment
shown in Fig. 1{\bf a} with applied field $H=3mT$ at and below 100K, namely $1.57\times10^{-10} Am^2$, the diamagnetic
moment in Fig. 1{\bf b} for field $H=30mT$ should be approximately $1.57\times10^{-9} Am^2$. Instead, it is
 three times larger, as indicated by the vertical line with double arrows in Fig. 1{\bf a}. Different measurements
 of the same superconducting sample should $not$ yield results that differ by a factor of 3 if they truly reflect properties of the 
 sample. (iii) the magnetic moment in Fig. 1{\bf a} increases rapidly when the temperature is lowered below 50K, which
 is $not$ the behavior expected for  the moment of a superconducting sample. These features call into question the interpretation that the
 results shown in Figs.   1{\bf a} and {\bf b}, published as Fig. 6 of Ref. \cite{e2022p},   reflect the physics of a superconducting sample  rather than experimental artifacts.

         \begin{figure} [t]
 \resizebox{8.5cm}{!}{\includegraphics[width=6cm]{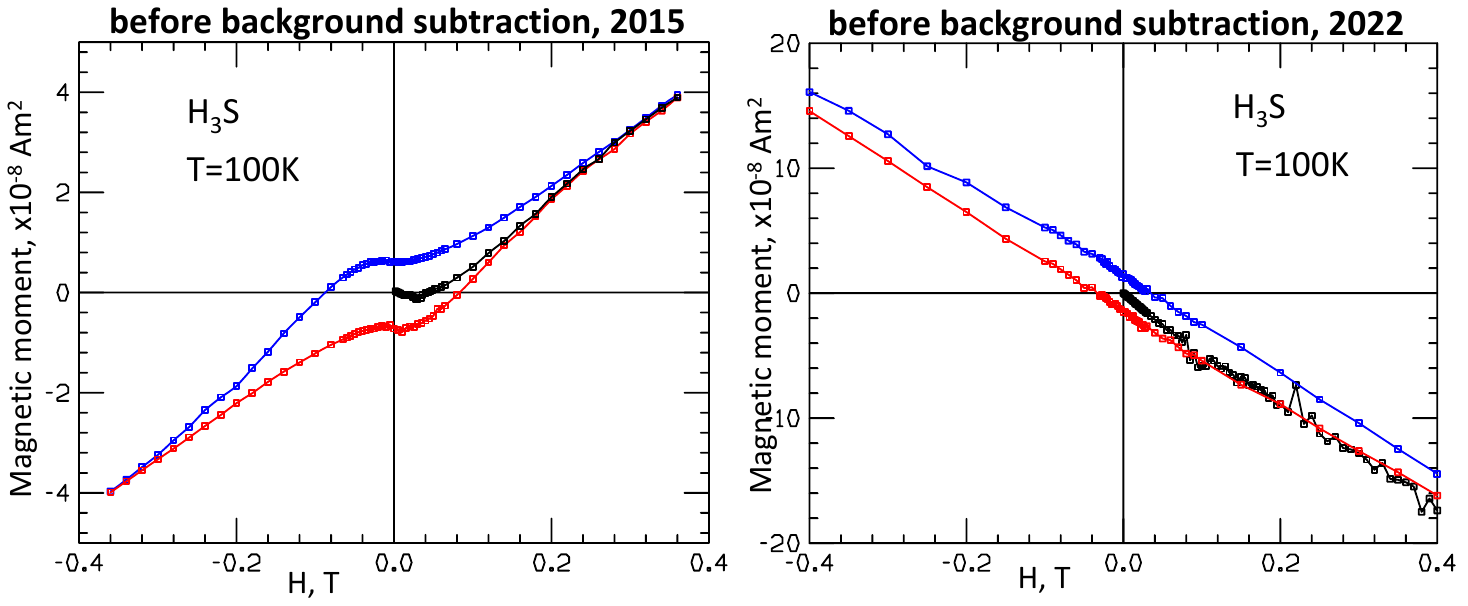}} 
 \caption { Raw data before background subtraction of 2015 experiments reported in Ref. \cite{e2022p} (left panel)
 versus raw data reported in 2022 in Ref. \cite{e2021p}  (right panel). Here as well as in Figs. 3, 4, 6 and 7,   black points  are the virgin curve,
 blue and red points correspond to the decreasing/increasing field branches of the hysteresis loop.}
 \label{figure1}
 \end{figure}

Moreover the  data shown in  Fig. 1{\bf b}, highlighted in Ref. \cite{stay} as proof of superconductivity, were superseded by new magnetization data on the same material (sulfur hydride) reported by the same group in 2022 \cite{e2021p}. In Ref. [6] of Ref. \cite{stay}, namely Ref. \cite{e2022p},  the new measurements are referred to as {\it ``the recent and more accurate measurements”}, and it is said
about them  {\it “only recently, we succeeded in improving our measurements of the magnetic susceptibility significantly”}. 
Ref. \cite{stay} is silent on those supposedly improved measurements. Here is why:

Figure 2 shows comparison
of hysteresis loops for the raw magnetic moment data for the 2015 experiments highlighted in Ref. \cite{stay} (left panel) versus the recently
measured data  \cite{e2021p}  at the same temperature (right panel). It can be seen that in the recently measured data
the response is strongly diamagnetic, of magnitude 5 times larger than the paramagnetic response of 2015. 
This large diamagnetic signal measured in 2022 can obviously not originate in a superconducting sample since (a) its magnitute is
far too large and (b) it persists above 200K \cite{e2021p}. Therefore, it must be a property of the background, i.e. the pressure cell.
This then nullifies the assertion of Ref. \cite{stay} that the tiny diamagnetic signal detected in the raw data
in 2015 {\it ``Even before background subtraction, the data for the virgin curve go negative,
in low applied fields, offering strong evidence that diamagnetism is observed in the raw data''} says anything relevant about
superconductivity of the sample itself.

          \begin{figure} [t]
 \resizebox{8.5cm}{!}{\includegraphics[width=6cm]{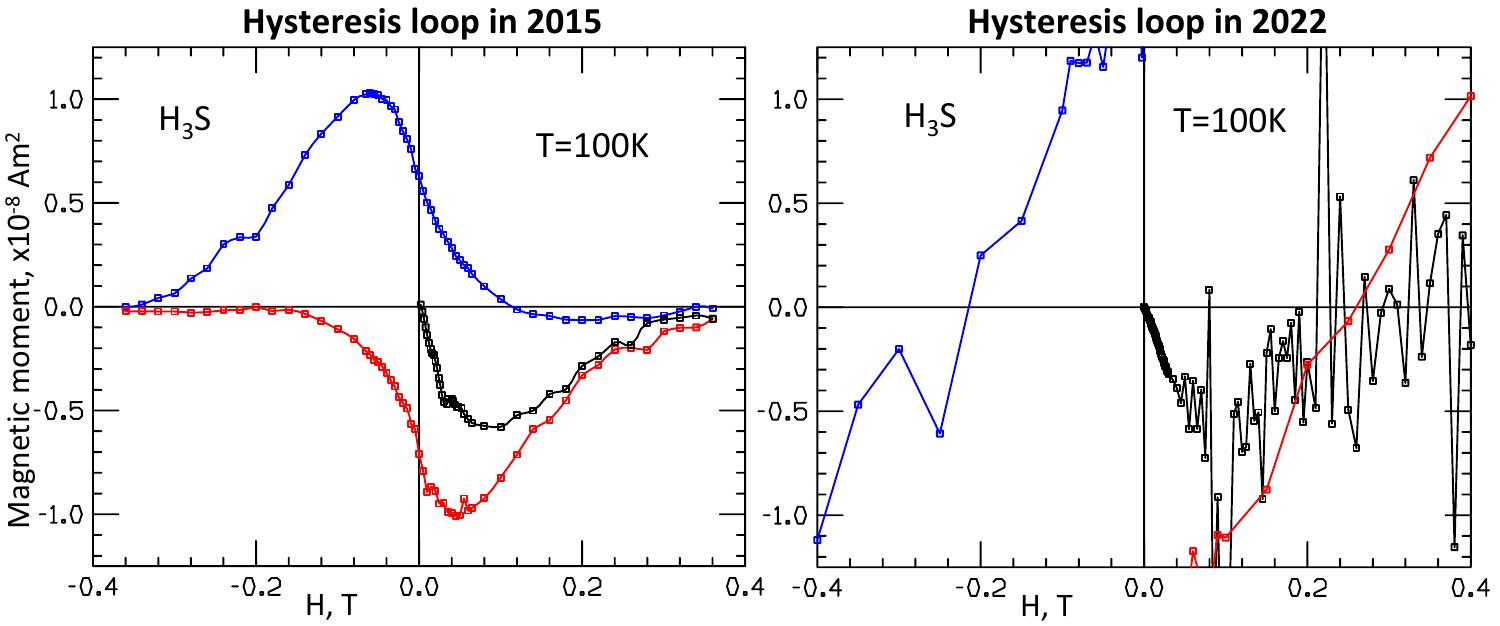}} 
 \caption { Hysteresis loops after background subtraction for 2015 data from Ref. \cite{e2022p}
 (left panel) versus the same for 2022 data from Ref. \cite{e2021p} (right panel).  }
 \label{figure1}
 \end{figure}
 
            \begin{figure} [t]
 \resizebox{8.5cm}{!}{\includegraphics[width=6cm]{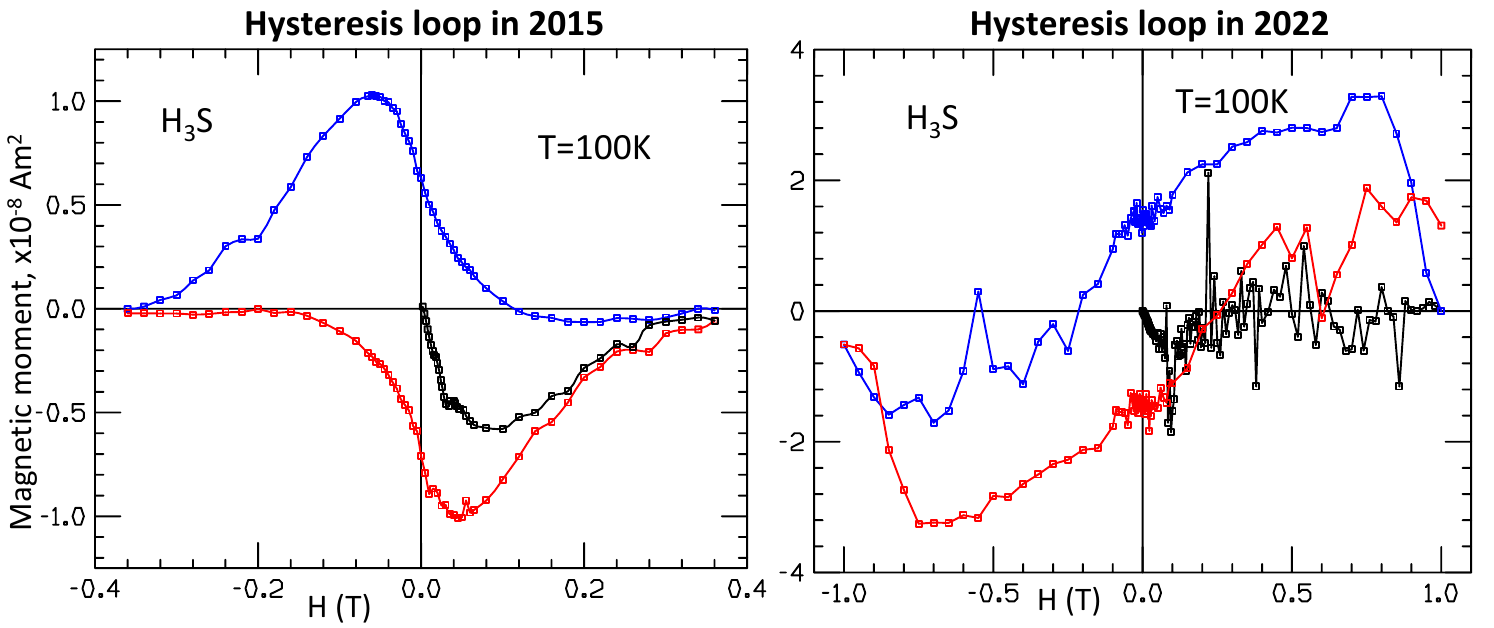}} 
 \caption {Hysteresis loops after background subtraction for 2015 data from Ref. \cite{e2022p}
 (left panel) versus the same for 2022 data from Ref. \cite{e2021p} for a wider range
 of magnetic field values (right panel). }
 \label{figure1}
 \end{figure}

In Fig. 3 I  compare   hysteresis loops measured in 2015 \cite{e2022p}  and 2022 \cite{reval}  at the same temperature after background subtraction, using
the same scales for horizontal and vertical axis. The samples were for the same material, $H_3S$, and
the dimensions of the samples were similar. Yet the figures show significant differences.
The 2015 figure indeed {\it ``has the main qualitative features expected of a superconducting hysteresis loop in the presence of flux trapping''} as remarked in Ref. \cite{stay}. The same cannot be said of the 2022 figure.
On the right panel it can be seen that the red data from the return part of the hysteresis  loop 
cross the virgin curve (black points) which is $not$ the expected behavior of a superconducting sample.
This is seen more clearly in Fig. 4 where   the range of H-values extends up to 1T for the 2022 data.

Moreover, when the recently published data on the diamagnetic response of hydrides under pressure \cite{e2021p}
are compared \cite{revalours} with the actually measured data as reported by the authors  \cite{reval},  it becomes apparent that
 the published data \cite{e2021p}  where $not$ derived from the measured data through scientifically accepted procedures \cite{arxivmine,revalours}. In other words, they were manipulated or worse. 
As an illustration, Fig. 5 shows comparison of what was published \cite{e2021p} (left panel) with what results
from what was reportedly measured \cite{reval}, as  analyzed  in Ref. \cite{revalours} (right panel).
Ref. \cite{revalours} identifies numerous instances where the actually measured data differ significantly 
with what was published, even for published data that were reported to be ``raw data'' \cite{e2021p,reval,revalours}.

             \begin{figure} []
 \resizebox{8.5cm}{!}{\includegraphics[width=6cm]{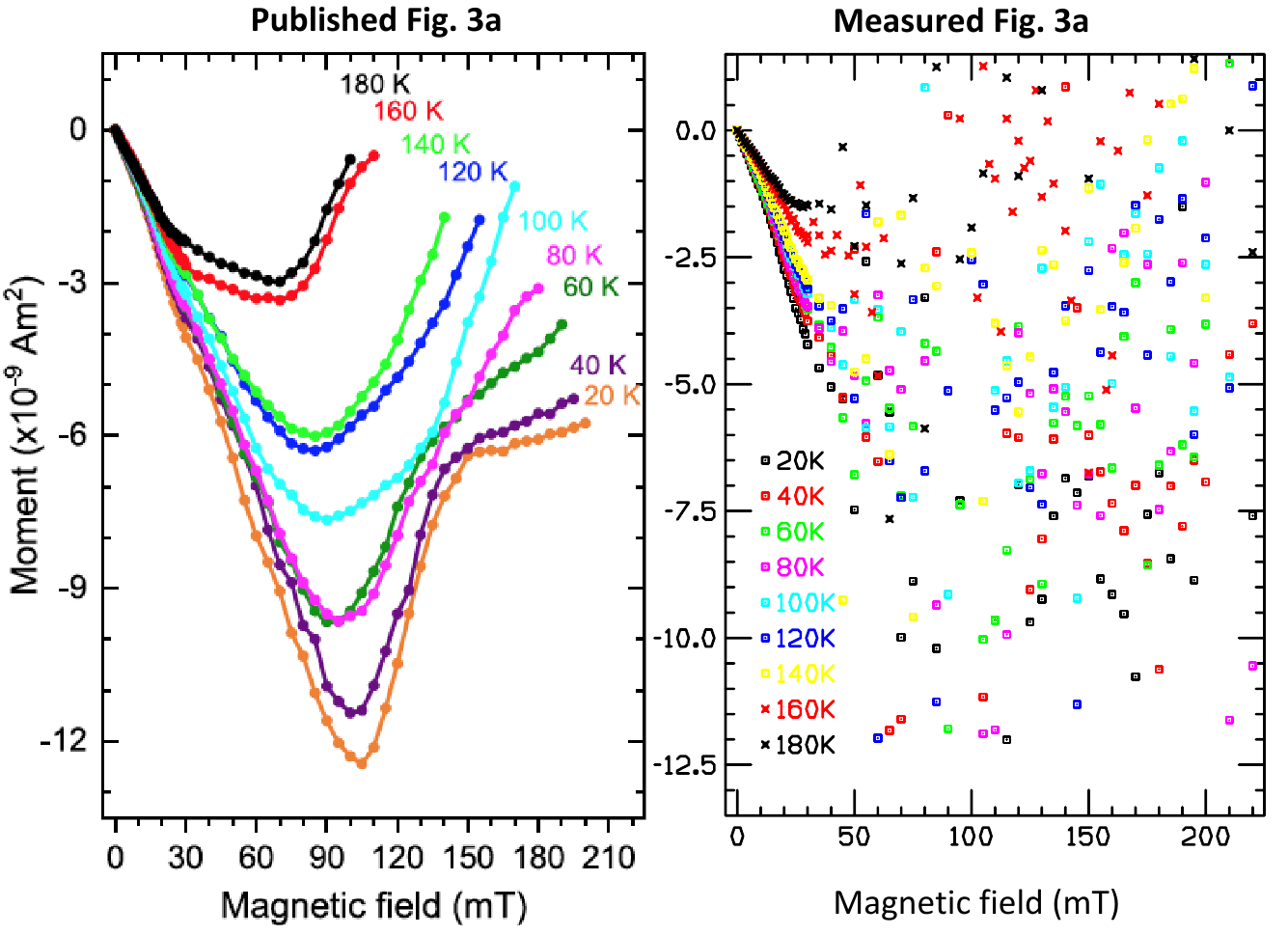}} 
 \caption {Comparison of what was published as depicting magnetic moment versus magnetic
 field of sulfur hydride after linear background subtraction  in 2022 \cite{e2021p}  (left panel) with  what was actually
 measured after the same background subtraction \cite{reval}, \cite{revalours}(right panel).}
 \label{figure1}
 \end{figure}

Given the stark difference between the left and right panels of Fig. 5, how can a reader know whether
there is another `panel'  associated with the right panel of Fig. 1 (Fig. 1{\bf e} of Ref. \cite{stay}), with results derived from actual raw data that could be as different from the right panel of Fig. 1 as the left and right panels of Fig. 5 are?
I have asked the authors of Ref. \cite{e2022p} repeatedly to make available the measured raw data underlying  their figure, they have
not responded.
 Therefore, whether or not the 10-year old magnetic data shown in Fig.  1 {\bf e} of Ref. \cite{stay}), that its 15 authors argue validate their opinion, are a true reflection of what was measured is unknown to them and to the rest of the world.

The only other magnetic data that the authors of Ref. \cite{stay} present to support their opinion are the hysteresis loops reported in the 
2015 Drozdov et al paper that started the field 10 years ago \cite{e2015}. An example is shown on the left panel of Fig. 6, compared to the hysteresis loop discussed earler (Fig. 1 right panel) that was 
also measured in 2015 but only published 7 years later \cite{e2022p}. We use the same scales for the horizontal and vertical axis for both panels.  The virgin curves are absent in the data published in 2015 \cite{e2015} for
unexplained reasons. The raw data underlying the 2015 published data are also  not available. 
The loops on the left and right panels of Fig. 6 look rather different, for unexplained reasons.
To understand whether or not they are consistent with the properties of a superconducting sample I subtract a linear background so that
the curves go to zero for large field, as was done in Ref. \cite{e2022p} to obtain the main panel of our Fig. 1 right panel.
The results are shown in Fig. 7. While the right panel can be said   {\it ``has the main qualitative features expected of
a superconducting hysteresis loop in the presence of flux trapping''}, as stated in Ref. \cite{stay}, the left panel {\it does not}.
As the field is decreased from its maximum value (blue curve) the magnetization should increase rather than decrease according to Faraday's law for 
a superconducting hysteresis loop in the presence of flux trapping

          \begin{figure} [t]
 \resizebox{8.5cm}{!}{\includegraphics[width=6cm]{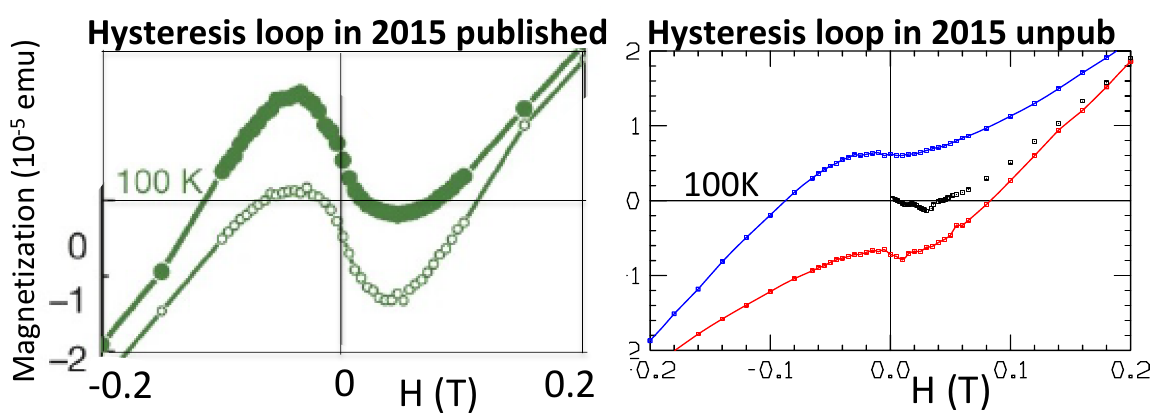}} 
 \caption {Hysteresis loops without background subtraction. Left panel: measured and published in 2015 \cite{e2015}.
 Lower (upper) curves correspond to increasing/decreasing field, the virgin curve is absent.
 Right panel:  same as inset of Fig. 1 right panel, measured in 2015, published in 2022 \cite{e2022p}. }
 \label{figure1}
 \end{figure}

In conclusion, I argue that  in view of all of the above the magnetic evidence for superconductivity in hydrides presented in Ref. \cite{stay} and analyzed here  {\it does not} provide strong evidence for superconductivity in these systems, contrary to the conclusion of Ref. \cite{stay}. In addition it should be noted that 
no control experiment has ever been reported showing that, under the same conditions of temperature and pressure with the same measuring apparatus
and pressure cell,  with a sample known to be non-superconducting or no sample, no hysteresis results from experimental artifacts or properties of the pressure cell. 
%In the absence of that information
% I argue that none of the magnetic information discssed in
%Ref. \cite{stay} is compelling evidence for superconductivity in these systems.

 We have questioned the validity of  other magnetic evidence argued to support the view that hydrides under high pressure exhibit superconductivity in other
 publications \cite{hmmre,dirkmre,persp}.

           \begin{figure} []
 \resizebox{8.5cm}{!}{\includegraphics[width=6cm]{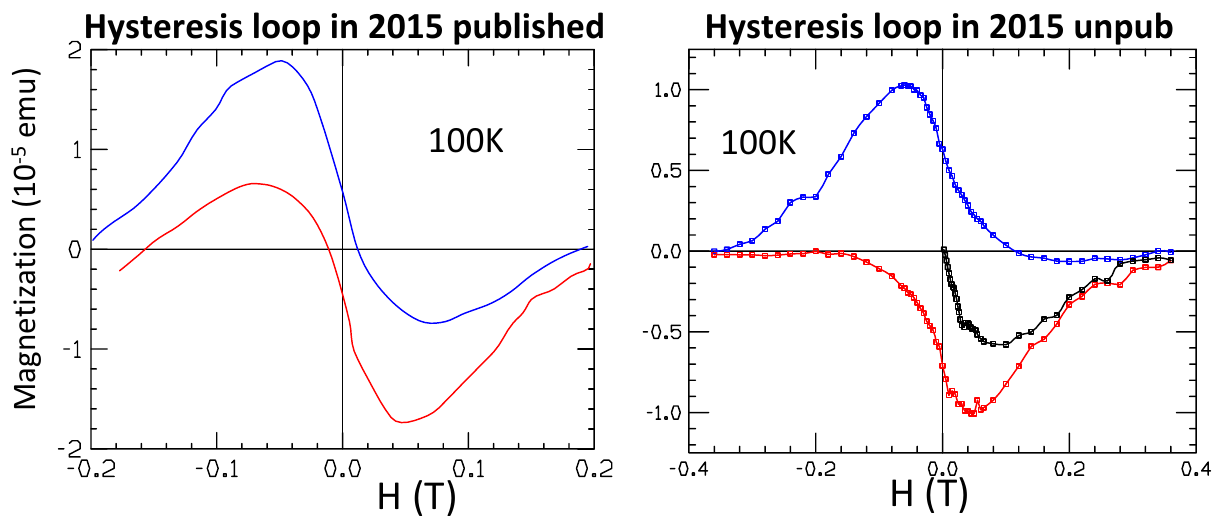}} 
 \caption {Same as Fig. 6 after linear background subtraction.  }
 \label{figure1}
 \end{figure}

The reality is that 10 years after it was heralded   what would have been a  momentous discovery deserving of a Nobel prize if true,  that ``conventional superconductivity'' \cite{e2015} eclipses the cuprates \cite{cuprates}
 as the highest $T_c$ superconductors, the field of hydride superconductivity is on life support. There is $no$ reliable reproducible magnetic evidence
of their claimed superconductivity. As time goes on,
 samples are not becoming more homogeneous nor better characterized, measurements are not becoming crisper and more representative of the physics
 of the materials studied, getting access to raw data is not becoming easier \cite{movie1,movie2}.
 Instead fraud \cite{roomt}, misrepresentation \cite{e2021p}, uncontrolled  one-shot slipshod non-reproduced experiments
 \cite{nrs,huang,timusk,struzhkin,550,nvcenter,mao2024}   are  the norm. None of the myriad of theoretical predictions \cite{lowp} of high $T_c$ superconductivity of hydrides or other light element
 materials at lower pressures, where samples could be larger and better characterized, have been realized \cite{aplmaterials}. 
 Yet the cheerleading continues unabated \cite{roadmap,stay,physics,nvcenter}. Why?

Because hydrides under pressure being superconducting would establish the validity of BCS electron-phonon theory to describe real superconductors,  in which the 15
superconductivity experts that authored Ref. \cite{stay}  firmly believe in, as does essentially the entire scientific community,
while hydrides under pressure not
being superconducting  would cast serious doubt on BCS’s validity to describe $any$ superconducting material.  In the field of hydride superconductivity the scientific method seems to have been thrown aside by
experimentalists and theorists alike grasping for straws that would show that BCS   theory is no longer impotent to predict any new superconducting material \cite{matthias}, rather than face the possibility that for almost 70 years   it has led everybody astray  and away from a unified description of superconductivity
for all materials  \cite{aplmaterials,xor,book}.

\begin{acknowledgements}
I am grateful to M. van Kampen for digitizing the data shown in Fig. 1 right panel,  from which the data shown in several   other Figs. were obtained,  and to F. Marsiglio, D. Van der Marel and M. van Kampen for collaboration
in topics discussed here.

\end{acknowledgements}

 \end{document}